\documentclass[aps,prb,twocolumn,showpacs,superscriptaddress, longbibliography, 10pt]{revtex4-1}

\usepackage{graphicx}
\usepackage{bm}
\usepackage{cleveref}
\usepackage{siunitx}
\usepackage{color}
\usepackage[normalem]{ulem}

\begin{document}

\title{Electrical switching of magnetic polarity in a multiferroic BiFeO$_3$ device at room temperature}

\author{N. Waterfield Price}
\email{noah.waterfieldprice@physics.ox.ac.uk}
\affiliation{Clarendon Laboratory, Department of Physics, University of Oxford, Parks Road, Oxford OX1 3PU, United Kingdom}
\affiliation{Diamond Light Source Ltd., Harwell Science and Innovation Campus, Didcot OX11 0DE, United Kingdom}

\author{R. D. Johnson}
\affiliation{Clarendon Laboratory, Department of Physics, University of Oxford, Parks Road, Oxford OX1 3PU, United Kingdom}
\affiliation{ISIS Facility, Rutherford Appleton Laboratory, Chilton, Didcot, OX11 0QX, United Kingdom}

\author{W. Saenrang}
\affiliation{Department of Materials Science and Engineering, University of Wisconsin-Madison, Madison, Wisconsin 53706, USA}

\author{A. Bombardi}
\affiliation{Diamond Light Source Ltd., Harwell Science and Innovation Campus, Didcot OX11 0DE, United Kingdom}

\author{F. P. Chmiel}
\affiliation{Clarendon Laboratory, Department of Physics, University of Oxford, Parks Road, Oxford OX1 3PU, United Kingdom}

\author{C.B. Eom}
\affiliation{Department of Materials Science and Engineering, University of Wisconsin-Madison, Madison, Wisconsin 53706, USA}

\author{P. G. Radaelli}
\affiliation{Clarendon Laboratory, Department of Physics, University of Oxford, Parks Road, Oxford OX1 3PU, United Kingdom}
\date{\today}

\begin{abstract}
We have directly imaged reversible electrical switching of the cycloidal rotation direction (magnetic polarity) in a $(111)_\mathrm{pc}$-BiFeO$_3$ epitaxial-film device at room temperature by non-resonant x-ray magnetic scattering.  Consistent with previous reports, fully relaxed $(111)_\mathrm{pc}$-BiFeO$_3$ epitaxial films consisting of a single ferroelectric domain were found to comprise a sub-micron-scale mosaic of magneto-elastic domains, all sharing a common direction of the magnetic polarity, which was found to switch reversibly upon reversal of the ferroelectric polarization without any measurable change of the magneto-elastic domain population.  A real-space polarimetry map of our device clearly distinguished between regions of the sample electrically addressed into the two magnetic states with a resolution of a few tens of micron.  Contrary to the general belief that the magneto-electric coupling in BiFeO$_3$ is weak, we find that electrical switching has a dramatic effect on the magnetic structure, with the magnetic moments rotating on average by 90$^{\circ}$ at every cycle.
\end{abstract}

\pacs{77.55.Nv, 75.85.+t, 61.05.cp}

\maketitle

\section{Introduction}

Multiferroicity as a phenomenon is defined by the presence of an antiferromagnetic order parameter that is \emph{quadratic} in the magnetic moments and is coupled \emph{linearly} to the ferroelectric (FE) polarization. Different multiferroic mechanisms differ in the nature of the magnetic order parameter: for the most common cycloidal multiferroics~\cite{mostovoy2006ferroelectricity}, this parameter is represented by a \emph{magnetic polarity}, and is commonly referred to as ``cycloidal rotation direction'' or ``magnetic chirality''. For the other two known varieties --- ferroaxial~\cite{johnson2011cu}  (including p-d hybridised~\cite{arima2007ferroelectricity}) multiferroics and exchange-striction multiferroics~\cite{chapon2004structural} --- this order parameter is represented by a magnetic helicity or a staggered exchange scalar field, respectively. Multiferroics are further classified into Type-I, where this order parameter is induced upon magnetic ordering in the presence of a pre-existing polarization, and Type-II, where, conversely, the polarization is induced by the appearance of this order parameter at a magnetic transition. One important consequence of the aforementioned linear coupling is that the magnetic polarity of a cycloid can be switched through reversal of the FE polarization by an electric field. Electrical control of magnetic domains at low temperature has indeed been demonstrated on bulk single crystals of several multiferroics by neutron polarimetry~\cite{radaelli2008electric, yamasaki2007electric, cabrera2009coupled} and magnetic x-ray scattering~\cite{fabrizi2010electric}. In the latter study, Fabrizi et al. were able to exploit the small x-ray beam size to image the gradual switching of magnetic cycloidal domains, a technique that has since been expanded to Type-I~\cite{johnson2013x} and helical Type-II~\cite{hiraoka2011spin} multiferroics.

Most of the multiferroic devices work carried out to date has employed the prototypical cycloidal Type-I multiferroic BiFeO$_3$ (BFO).  In this material, the direction of electric polarization can be switched within, and between, ferroelastic domains that are equivalent by pseudo-cubic symmetry~\cite{zavaliche2006multiferroic}. The direction of the cycloidal propagation vector~\cite{lee2008single} and the spin rotation plane~\cite{zhao2006electrical, lebeugle2008electric, lee2008single} can also be manipulated, since they are coupled to the crystal via magneto-elastic strain~\cite{sando2013crafting}. In the attempt to create a prototypical memory device, these switching mechanisms have been exploited to control the magnetization of a thin ferromagnetic overlayer~\cite{chu2008electric, bea2008mechanisms, lebeugle2010exchange, heron2011electric, heron2014deterministic, sando2014bifeo3, gao2016electric}. In fact, it has been shown that the overlayer magnetization can be completely reversed upon switching between 71$^\circ$ ferroelastic domains~\cite{heron2011electric, heron2014deterministic} --- an effect, which was believed to be forbidden by simple symmetry considerations.  In spite of these encouraging results, there has been relatively little progress in relating macroscopic device switching to microscopic changes in the magnetic structure at the atomic level, and in particular to the expected reversal of the magnetic polarity.  Obtaining such information in an epitaxial film device requires a polarimetric scattering technique with both high sensitivity and high spatial resolution --- a combination that had not been achieved so far.

In this article, we present the results of a non-resonant x-ray magnetic scattering (NXMS) study in which we have directly demonstrated electrical reversal of magnetic polarity in a BFO epitaxial-film device at room temperature. In our device architecture, the full component of the electric polarization is aligned parallel to the applied (surface-normal) electric field, such that 180$^\circ$ switching of both electric polarization and magnetic polarity may be achieved. By combining circularly polarized x-rays and post-scatter polarization analysis, we determined the magnetic polarity of the cycloidal magnetic structure for two FE polarization states. We further use this to map the magnetic polarity over the device with a $\SI{50}{\micro\metre} \times \SI{50}{\micro\metre}$ resolution, demonstrating that the magnetic polarity of micron scale regions of the sample may be addressed by an electric field --- an important result in the quest for practical applications of this class of materials.

\begin{figure}[h!]
\includegraphics[width=0.47\textwidth]{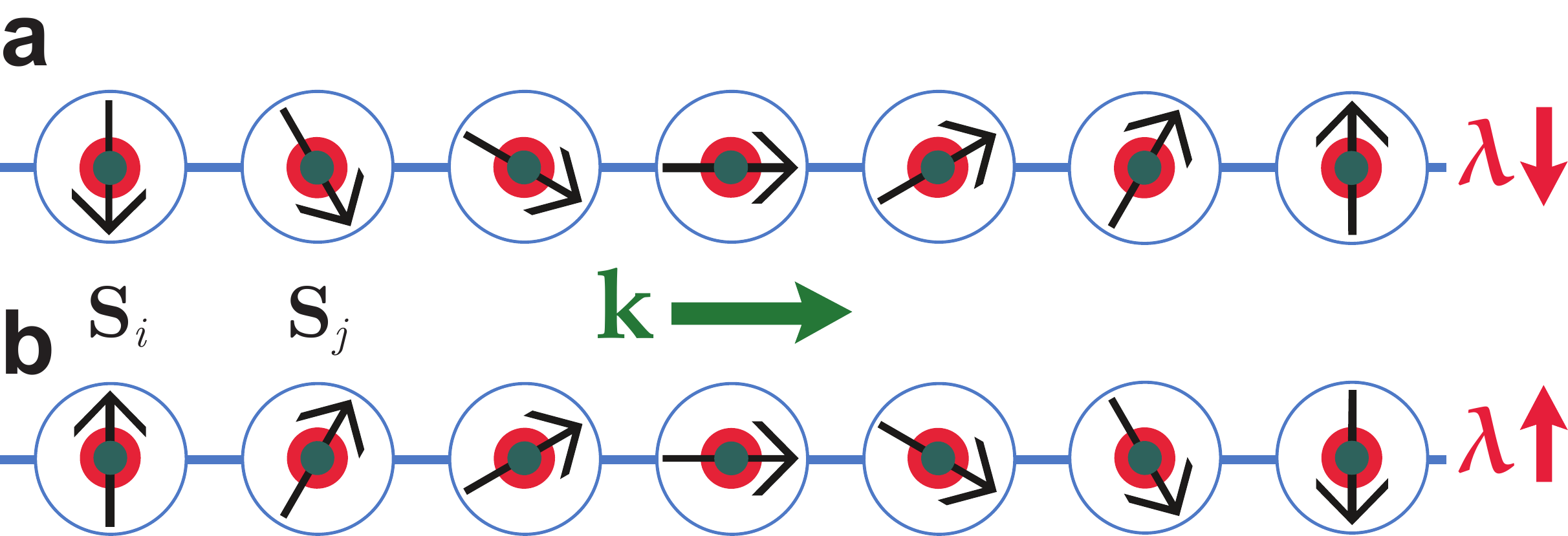}
\caption{\label{fig:magpol} Magnetic Polarity. \textbf{a,b,} Magnetic spin cycloids of opposite magnetic polarity.}
\end{figure}

The spontaneous electric polarization, $\mathbf{P}$, is directed along one of the $[111]_\mathrm{pc}$ axes of the BFO pseudo-cubic (pc) cell and is associated with a ferroelastic distortion along the same direction, resulting in a rhombohedrally distorted perovskite structure. It has been shown that the polarization can be switched by \SI{71}{\degree}, \SI{109}{\degree}, or \SI{180}{\degree} between the 8 possible $[111]_{pc}$ directions, where switching by \SI{71}{\degree}, \SI{109}{\degree} is also accompanied by a switching of the ferroelastic state~\cite{zavaliche2006multiferroic}. Since the rhombohedral axis is unchanged when the polarization is switched by \SI{180}{\degree}, the ferroelastic domain remains unchanged. The magnetic structure of BFO can be described locally as G-type but the spin-orbit interaction in the presence of the FE polarization drives an incommensurate cycloidal modulation of the spins (the direct Dzyaloshinksii-Moriya effect~\cite{kadomtseva2004space}), rotating them in a plane containing  $\mathbf{P}$. The propagation vector of the cycloid, $\mathbf{k}$, is always orthogonal to $\mathbf{P}$, and can take one of three symmetry equivalent directions in the rhombohedral lattice: $\mathbf{k}_1 = (\delta, \delta, 0)_\mathrm{h}, \mathbf{k}_2 = (\delta, -2 \delta, 0)_\mathrm{h}, \mathbf{k}_3 = (-2 \delta, \delta, 0)_\mathrm{h}$. Here the subscript h denotes the hexagonal setting of the rhombohedral unit cell, and $\delta = 0.0045$ at \SI{300}{\kelvin}. In the case of a cycloidal magnetic structure, the quadratic order parameter is the magnetic polarity,  defined as $\boldsymbol{\lambda} = \mathbf{k} \times \mathbf{S}_i \times \mathbf{S}_j$ where $\mathbf{S}_i$ and $\mathbf{S}_j$ are spins on adjacent sites (sequential in the direction of $\mathbf{k}$), as depicted in \Cref{fig:magpol}. This appears in a term in the Landau free energy expansion of the form $\propto \boldsymbol{\lambda} \cdot \mathbf{P}$, which couples the FE polarization to the magnetic structure.

\begin{figure}[h]
\includegraphics[width=0.47\textwidth]{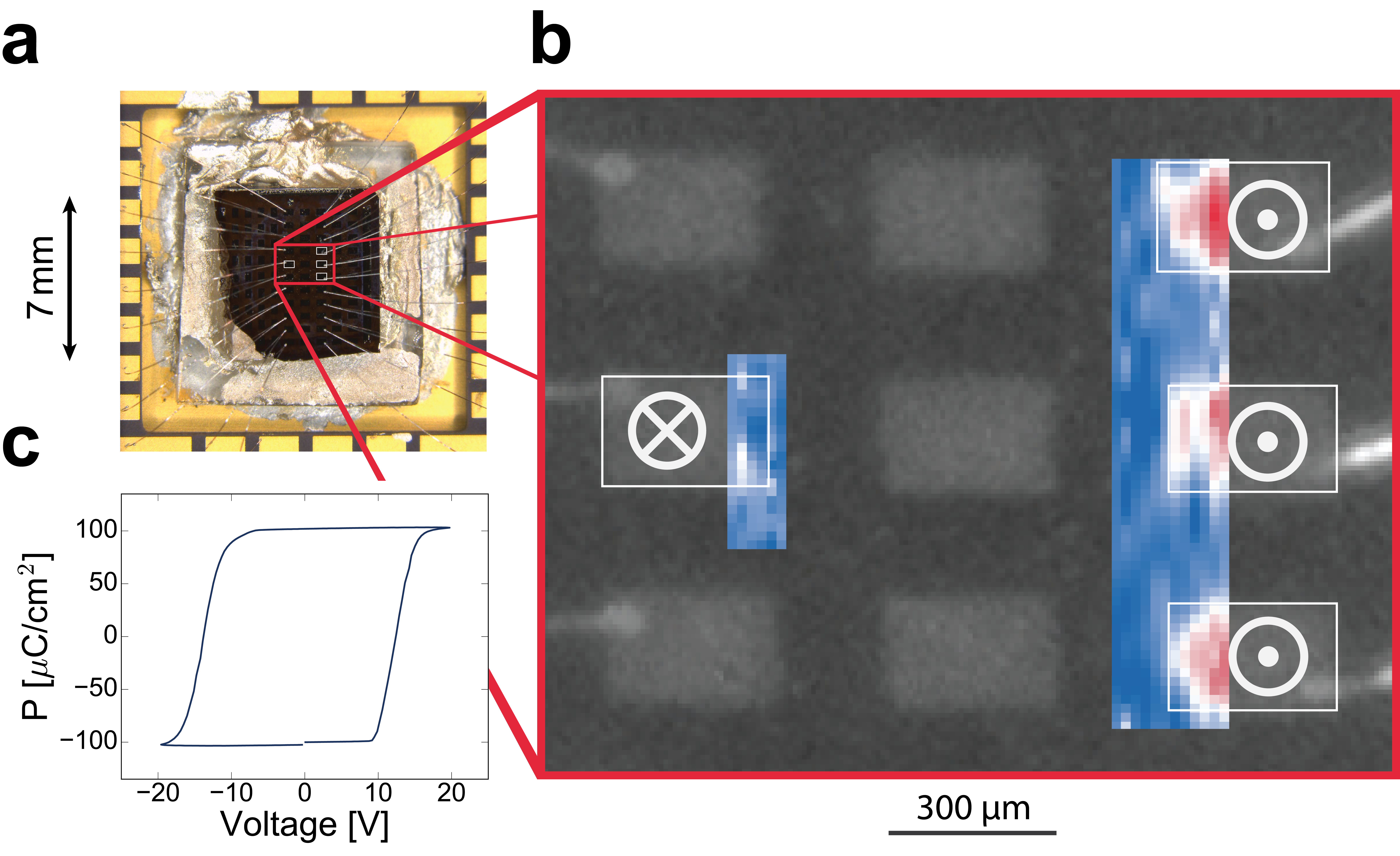}
\caption{\label{fig:char} $(111)_\mathrm{pc}$-BFO film device. \textbf{a,} Photo of the $(111)_\mathrm{pc}$-BFO film. \textbf{b,} Optical image of the $(111)_\mathrm{pc}$-BFO film surface. Maps of the magnetic polarity (see \cref{fig:real_space_maps}) are overlaid showing regions of the sample switched into the FE\,$\downarrow$ and FE\,$\uparrow$ state and the direction of the FE polarization is shown by the white symbols. \textbf{c,} P-E hysteresis loop measurement. }
\end{figure}

\begin{figure}[b]
\includegraphics[width=0.47\textwidth]{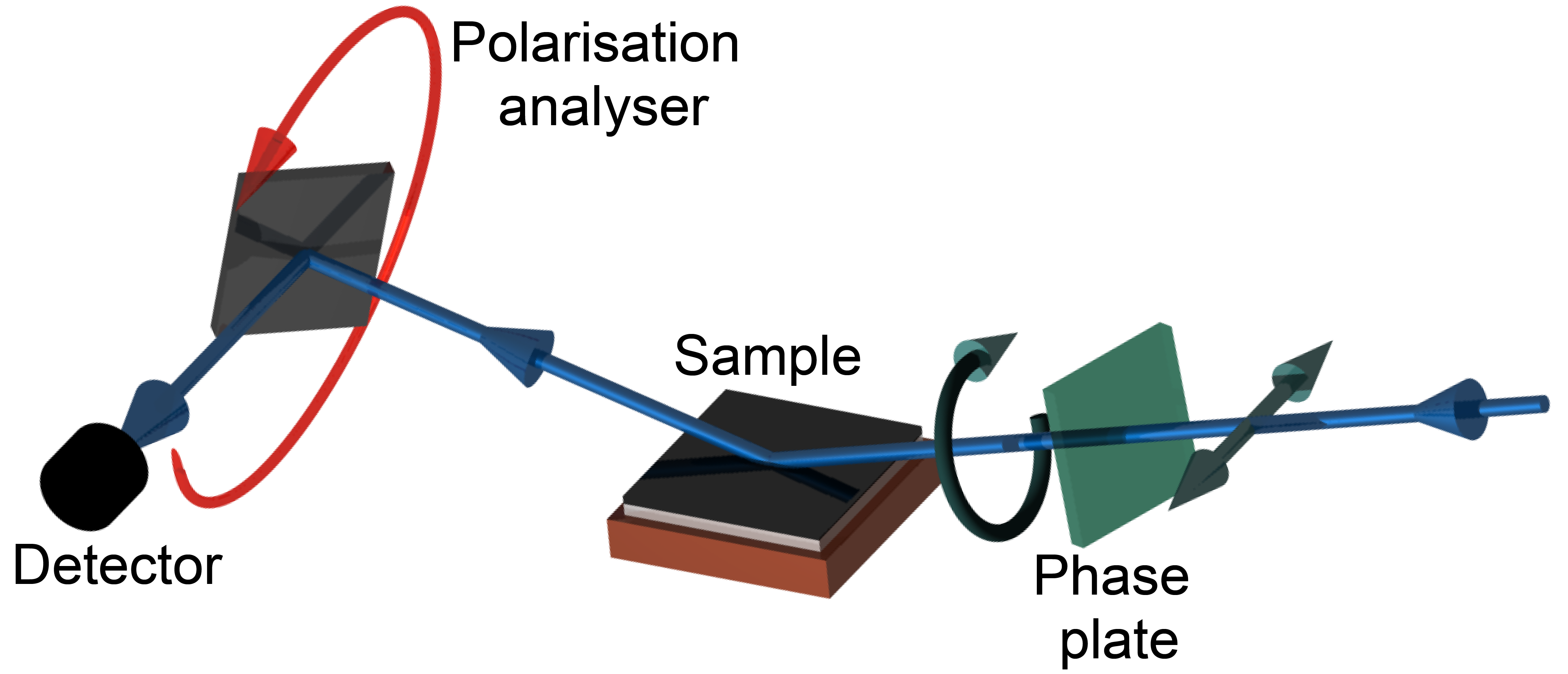}
\caption{\label{fig:expsetup} Non-resonant magnetic x-ray scattering experimental setup. The incident and scattered x-ray beam directions are indicated by the blue arrows. The incident x-rays are linearly polarized with their $\mathbf{E}$-field vector perpendicular to the scattering plane ($\sigma$-polarized). These may be converted to circularly polarised x-rays of either handedness using a diamond phase plate. The polarization of the scattered x-rays is measured using a graphite analyser crystal which can be rotated to select any linear polarization channel.}
\end{figure}

\begin{figure*}[t]
\includegraphics[width=\textwidth]{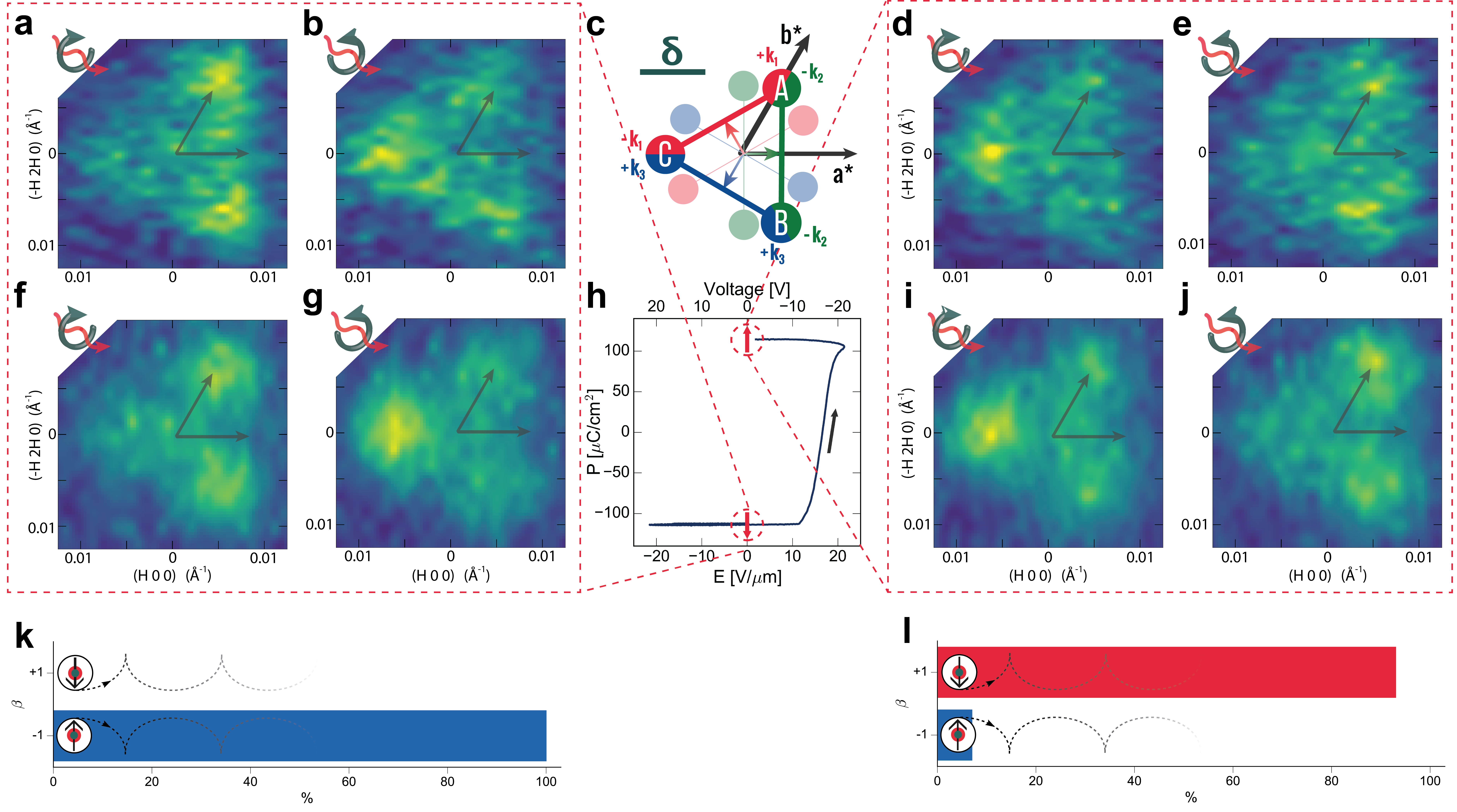}
\caption{\label{fig:rsms} Magnetic diffraction reciprocal space maps. \textbf{a,b,}  NXMS Reciprocal space maps about $(009)_\mathrm{h}$ with left circular and right circular polarized light, respectively in the virgin FE\,$\downarrow$ state. \textbf{f,g,} Fitted NXMS reciprocal space maps about $(009)_\mathrm{h}$ with left circular and right circular polarized light, respectively in the virgin, FE\,$\downarrow$ state. \textbf{d,e,}  NXMS Reciprocal space maps about $(009)_\mathrm{h,}$ with left circular and right circular polarized light, respectively in the switched (nominally) FE\,$\uparrow$ state. \textbf{i,j,} Fitted NXMS reciprocal space maps about  $(009)_\mathrm{h}$ with left circular and right circular polarized light, respectively in the switched (nominally) FE\,$\uparrow$ state. \textbf{c,} Schematic showing the effect of the monoclinic distortion on the diffracted signal, showing the undistorted (translucent) and distorted (opaque) diffraction patterns. \textbf{h,} FE polarization of the sample measured while switching into the FE\,$\uparrow$ state. The points at which the reciprocal space maps in \textbf{a,b} and \textbf{d,e} were measured are labelled with $\downarrow$ and $\uparrow$, respectively. The black arrow indicates the electric-field sweep direction. \textbf{k,l,} The percentage fraction of both the $\beta = +1$ and $\beta = -1$ magnetic polarities of the measured regions. The scale bar in \textbf{c} shows the magnitude of the propagation vector (all reciprocal space maps on the same scale) and the reciprocal lattice directions (in the hexagonal setting) are indicated by the tranlucent black arrows on the reciprocal space maps. All measurements were taken at room temperature.} \vspace*{-0.5cm}
\end{figure*}

\section{Experimental}

\SI{1}{\micro\metre} thick epitaxial films of $(111)_\mathrm{pc}$ BFO were grown by double gun off-axis sputtering onto a $(111)_\mathrm{pc}$ surface normal SrTiO$_3$ single crystal substrate~\cite{das2006synthesis}. A \SI{30}{\nano\metre} thick SrRuO$_3$ layer was first deposited on the SrTiO$_3$ substrate by \SI{90}{\degree} off-axis sputtering before the BFO was grown, which serves as a bottom electrode~\cite{eom1992single}. \SI{300}{\micro\metre} x \SI{200}{\micro\metre}  x \SI{25}{\nano\metre} Pt top electrodes were patterned on the surface of the film using photolithography, which were then wirebonded to a chip carrier [see \cref{fig:char}(a, b)]. This setup allows electric fields to be applied along the $(111)_\mathrm{pc}$ direction, perpendicular to the surface of the film. As shown in \cref{fig:char}(c) The films display excellent FE characteristics with a remnant polarization along the $[111]_\mathrm{pc}$ direction of $\mathbf{P} = \SI{102}{\micro\coulomb/\cm^2}$, comparable to the highest reported literature values for $(111)_\mathrm{pc}$ oriented BFO films~\cite{das2006synthesis, li2004dramatically, bai2005destruction}. We label the FE domain with polarization directions along $[111]_\mathrm{pc}$ (surface normal, pointing out of the film) and $[\bar{1}\bar{1}\bar{1}]_\mathrm{pc}$ (surface normal, pointing into the film) as FE\,$\uparrow$ and FE\,$\downarrow$, respectively. The as-grown state of the film is a FE\,$\downarrow$ monodomain, as determined by piezoresponse force microscopy on other representative samples. Regions of the sample were switched into the FE\,$\uparrow$ polarization state prior to the experiment by applying a number of successive triangular waves (see Supplementary Material S-I~\cite{Suppl}) at \SI{5}{\kilo\hertz}, with a maximum voltage of \SI{22}{\volt}. To test the reversibility of this process, we switched some of these regions back into the FE\,$\downarrow$ polarization state, using an inverted signal.

The magnetic state of the BFO film was probed by NXMS, which can measure the direction of propagation and periodicity of the spin cycloid, as well as the spin rotation plane. Using x-ray polarimetry, the magnetic polarity of the cycloidal magnetic structure may also be determined. The synchrotron NXMS experiments were performed on beamline I16 at Diamond Light Source (UK) {using a six-circle kappa diffractometer in the reflection (vertical scattering) geometry, as shown in \cref{fig:expsetup}. The incident x-ray beam energy was tuned to \SI{4.8}{\kilo\electronvolt}, which was selected for the following reasons: a) It was off-resonance of all chemical elements present in the sample. b) Absorption both by air and the Pt electrodes on the surface of the device was negligible. c) The probability of multiple scattering processes was reduced. Conversion of the incident x-ray polarization from $\sigma$-polarized x-rays to circularly polarized x-rays was achieved using a \SI{100}{\micro\meter}-thick diamond quarter-wave plate. The scattered x-ray polarization was determined using a pyrolitic graphite polarization analyser crystal scattering at the (004) reflection, built on the detector arm of the diffractometer such that any linear polarization channel may be selected. Measurements were taken at an azimuthal angle of $\phi = \SI{94}{\degree}$ with respect to the $(100)_h$ direction, identified as an experimental geometry in which multiple scattering was minimized.

\begin{figure*}[ht]
\includegraphics[width=\textwidth]{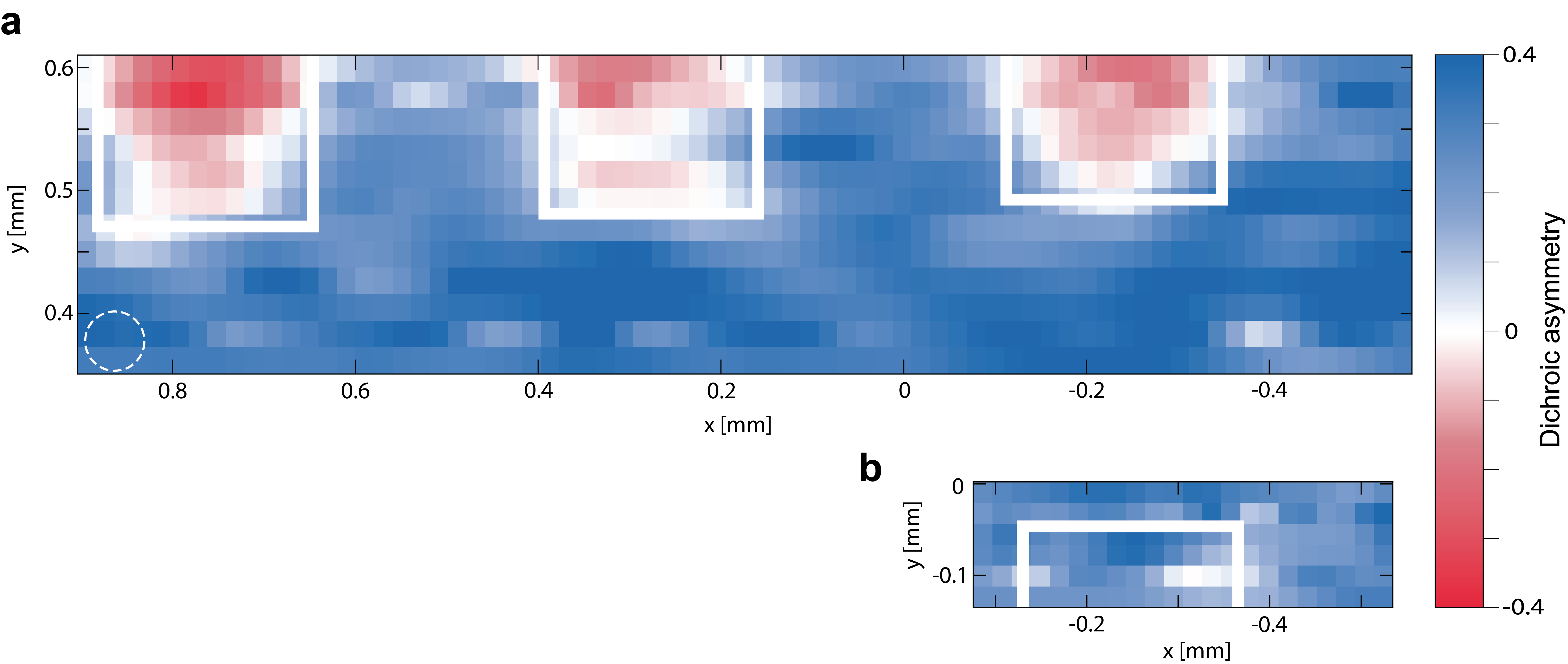}
\caption{\label{fig:real_space_maps} Real space maps. \textbf{a,} Magnetic polarity real space map of an area of the sample containing regions that have been switched into the FE\,$\uparrow$ state. The white dotted circle in the lower left corner indicates the spatial resolution of the measurement. \textbf{b,} Magnetic polarity real space map of an area of the sample containing regions that have been switched into the FE\,$\uparrow$ state and then switched back into the FE\,$\downarrow$ state. The regions underlying the Pt electrodes are outlined by the white solid lines and the surrounding regions are in the virgin FE\,$\downarrow$ state. The scale and color bars refer to both (a) and (b). Based upon the characterisation of switched and unswitched regions shown in Figure \ref{fig:rsms}, a dichroic asymmetry of $+0.4$ and $-0.4$ corresponds to a magnetic polarity of $\beta=-1$ and $\beta=+1$, respectively. $x$ and $y$ are two orthogonal directions that describe the laterial sample translation. These maps have been overlayed on an image of the sample in \cref{fig:char}(c).}
\end{figure*}

To determine cycloidal magnetic polarity, we measured by NXMS the magnetic satellite reflections which occur near the structurally-forbidden $\mathbf{N}=(009)_\mathrm{h}$ Bragg peak~\cite{waterfieldprice2016coherent} using circularly polarized x-rays. For a single magnetic domain with a propagation vector $\mathbf{k}_i$, the expected diffraction pattern is a pair of reflections located at $\mathbf{N} \pm \mathbf{k}_i$. As shown in a previous x-ray magnetic linear dichroism photo-electron emission microscopy (XMLD-PEEM) study of an an identical $(111)_\mathrm{pc}$-BFO film~\cite{waterfieldprice2016coherent}, the typical domain size is $\sim\SI{100}{\nano\metre}$, thus approximately equal populations of all three \textbf{k}-domains fall under the x-ray spot. In this case, one would expect to observe three pairs of peaks, as depicted by the lightly shaded circles in \cref{fig:rsms}(c). However, each magnetic domain is accompanied by a small monoclinic distortion, tilting the $ab$-plane in a direction orthogonal to $\mathbf{k}$. This moves the positions of the diffraction peaks such that they overlap, so that, instead of a star of six satellite peaks, only three composite peaks are observed~\cite{waterfieldprice2016coherent} as schematically shown in \cref{fig:rsms}(c).

\section{Results and discussion}

For a given handedness of the incident beam polarization, the diffracted signal of the magnetic satellites is highly sensitive to the magnetic polarity of the magnetic domains illuminated by the x-ray spot. \Cref{fig:rsms}(a) and (b) show, for a region of the sample in the `virgin' FE\,$\downarrow$ state, reciprocal space maps of the magnetic satellites about the $\mathbf{N}=(009)_\mathrm{h}$ position for left circular and right circular polarized light, respectively. \Cref{fig:rsms}(d) and (e) show the equivalent measurement on a region of the sample that has been switched into the FE\,$\uparrow$ state. We note that, as the diffraction intensity is centred at the same point in reciprocal space for the FE\,$\downarrow$ and FE\,$\uparrow$ states, both states are in the same pseudo-rhombohedral ferroelastic domain. The data agrees well with the corresponding simulations, shown in \cref{fig:rsms}(f,g) and \cref{fig:rsms}(i,j), respectively, and with previous measurements taken with higher statistics~\cite{waterfieldprice2016coherent}. The simulations shown here are generated with the parameters obtained by simultaneously fitting all reciprocal space maps, with the population of the two magnetic polarity states allowed to freely refine, along with a global depolarization factor (see Supplemental Material S-II~\cite{Suppl}). Here, we have assumed equal populations of the three $\textbf{k}$-domains, which, for this direction of the incident polarization, leads to equal intensities for the left-hand pair of peaks (labeled  $\mathrm{A}$ and $\mathrm{B}$ in \cref{fig:rsms}(c)). A theoretical intensity calculation for the composite magnetic peaks shown in \cref{fig:rsms} yields
\begin{equation}
  \frac{I_\mathrm{C}}{I_{\mathrm{A}, \mathrm{B}}} \approx \frac{1 + \beta\gamma}{1 - \beta\gamma/2}
\end{equation}
where the subscript denotes the individual peaks, labelled as  $\mathrm{A}$, $\mathrm{B}$, and $\mathrm{C}$ (see \cref{fig:rsms}(c)) and
where $\gamma = +1$ or $-1$ for left circular and right circular polarized light, respectively. The coupling between the magnetic polarity and FE polarization is parametrized by $\beta = - \frac{\boldsymbol{\lambda} \cdot \hat{\mathbf{n}}}{|\boldsymbol{\lambda}|}$, where $\hat{n}$ is the film surface normal. In agreement with a previous study on a bulk single crystal sample~\cite{johnson2013x}, we find that magnetic polarity of a given domain is aligned antiparallel to the FE polarization, hence $\beta = +1$ or $-1$ for the fully-polarized FE\,$\uparrow$ state and FE\,$\downarrow$ state, respectively.

As shown in \cref{fig:rsms}(k,l), the simultaneous fit of the reciprocal space maps yielded a \SI{93.0(6)}{\percent} switch of magnetic polarity from the virgin FE\,$\downarrow$ state to the FE\,$\uparrow$ state. The depolarization factor was found to be 0.194(4) (see Supplemental Material S-II~\cite{Suppl}), where this accounts for the imperfect circular polarization of the x-ray beam and contributions from any multiple scattering. The unswitched \SI{7}{\percent} could be explained by a fraction of the film being pinned in the FE$\downarrow$ state, likely due to the bias at the BFO/SrRuO$_3$ interface that stabilises a FE↓ monodomain in the as-grown BFO film~\cite{fong2006stabilization,giencke2014tailoring}. A small pinned fraction would not be detected in the FE polarisation measurement in \cref{fig:rsms}(h) since this measurement is relative i.e. there is an arbitrary offset in the magnitude of $\mathbf{P}$. Furthermore, although the majority of the beam intensity was focussed into an $\approx$ \SI{50}{\micro\metre} x \SI{50}{\micro\metre} footprint, low intensity tails of the beam will extend slightly beyond this so it is possible that a small portion of the surrounding film in the FE$\downarrow$ state was illuminated.

The response of the magnetic structure to electrical switching was further investigated by mapping in real space the magnetic polarity over an extended region of the device. \Cref{fig:real_space_maps}(a) and (b) show real space maps of an area of the sample containing regions that have been switched up and an area containing a region that has been switched up and then down again, respectively (see \cref{fig:char}(c)). These images were collected by rastering a \SI{50}{\micro\metre} x \SI{50}{\micro\metre} beam over the sample surface and recording the intensity of peak C at each point. The intensity was measured with both left-handed (lc) and right-handed (rc) circular polarization at every position. The maps were constructed by plotting the dichroic asymmetry, $(I_\mathrm{lc} - I_\mathrm{rc})/(I_\mathrm{lc} + I_\mathrm{rc})$. In \Cref{fig:real_space_maps}(a) it is clear that, for the majority of the sample region which has been switched into the FE$\uparrow$ state, the magnetic polarity has also switched relative to the unswitched film in the FE$\downarrow$ state. The areas where the magnetic polarity appears not to have switched are concentrated around the edges of the switched electrodes. This can largely be explained by the resolution limitation of the measurement (indicated by the white dotted circle in \cref{fig:real_space_maps}(a)), but it is also possible that electric-field edge effects resulted in incomplete FE switching at the outside of the pads. By contrast, the sample region below the electrical pad that has been switched into the FE$\uparrow$ state and then \emph{back} into the FE$\downarrow$ state, has unchanged magnetic polarity relative to the surrounding region of unswitched BFO, indicating that the magnetic state is recovered entirely.

\section{Conclusion}
In summary, we have demonstrated electrical switching of magnetic polarity in a BFO device at room-temperature, a direct measurement of multiferroic coupling. The observed reversal of the magnetic polarity of the cycloid represents a major rearrangement of the magnetic structure, with magnetic moments rotating on average by \SI{90}{\degree} upon switching.  This is in contrast to the general belief that the coupling between magnetism and ferroelectricity is rather weak in Type-I multiferroics~\cite{khomskii2009trend} but is in fact in full agreement with theoretical predictions~\cite{kadomtseva2004space}. Our experimental demonstration of polarity switching in a BFO device represents a crucial step towards developing full control of the interfacial exchange interactions in BFO-based composite multiferroic devices.

\section*{Acknowledgements}
We acknowledge Diamond Light Source for time on Beam line I16 under Proposal MT12837-1. The work done at the University of Oxford (N.W.P., R.D.J., F.P.C., and P.G.R.) was funded by EPSRC grant No. EP/M020517/1, entitled “Oxford Quantum Materials Platform Grant.” The work at University of Wisconsin-Madison (C.B.E. and W.S.) was supported by the Army Research Office through grant W911NF-13-1-0486. R. D. J. acknowledges support from a Royal Society University Research Fellowship.

\bibliography{switching_paper_references}

\end{document}